\documentclass[%
 reprint,,bibtex,
superscriptaddress,
nofootinbib,
bibnotes,
 amsmath,amssymb,
 aps,
prl,
]{revtex4-1}
\usepackage{float}
\usepackage[dvipsnames]{xcolor}
\usepackage{pdfpages}
\usepackage{units}
\usepackage{graphicx}
\usepackage{dcolumn}
\usepackage{bm}
\usepackage{hyperref}
\usepackage{tabularx}
\usepackage[utf8]{inputenc}

\newcommand{\ub}{\mu_B\unit[]{/u.c}}
\begin{document}

\preprint{APS/123-QED}

\title{Proximity Driven Enhanced Magnetic Order at Ferromagnetic Insulator / Magnetic Topological Insulator Interface
}

\author{Mingda Li}
\email{mingda@mit.edu}
\affiliation{Department of Nuclear Science and Engineering, Massachusetts Institute of Technology, Cambridge, MA 02139, USA}
\affiliation{Fracsis Bitter Magnet Lab, Massachusetts Institute of Technology, Cambridge, MA 02139, USA}
\affiliation{Condensed Matter Physics and Materials Science Department, Brookhaven National Laboratory, Upton, New York 11973, USA}

\author{Cui-Zu Chang}
\email{czchang@mit.edu}
\affiliation{Fracsis Bitter Magnet Lab, Massachusetts Institute of Technology, Cambridge, MA 02139, USA}

\author{Brian. J. Kirby}
\affiliation{Center for Neutron Research, National Institute of Standards and Technology, Gaithersburg, Maryland 20899, USA}

\author{Michelle Jamer}%
\affiliation{Department of Physics, Northeastern University, Boston, MA 02115, USA}

\author{Wenping Cui}
\affiliation{Department of Physics, Boston College, Chestnut Hill, MA 02467, USA}

\author{Lijun Wu}
\affiliation{Condensed Matter Physics and Materials Science Department, Brookhaven National Laboratory, Upton, New York 11973, USA}

\author{Peng Wei}
\affiliation{Fracsis Bitter Magnet Lab, Massachusetts Institute of Technology, Cambridge, MA 02139, USA}

\author{Ferhat Katmis}
\affiliation{Fracsis Bitter Magnet Lab, Massachusetts Institute of Technology, Cambridge, MA 02139, USA}

\author{Yimei Zhu}
\affiliation{Condensed Matter Physics and Materials Science Department, Brookhaven National Laboratory, Upton, New York 11973, USA}

\author{Don Heiman}
\affiliation{Department of Physics, Northeastern University, Boston, MA 02115, USA}

\author{Ju Li}
\affiliation{Department of Nuclear Science and Engineering, Massachusetts Institute of Technology, Cambridge, MA 02139, USA}
\affiliation{Department of Material Science and Engineering, Massachusetts Institute of Technology, Cambridge, MA 02139, USA}

\author{Jagadeesh S. Moodera}
\email{moodera@mit.edu}
\affiliation{Fracsis Bitter Magnet Lab, Massachusetts Institute of Technology, Cambridge, MA 02139, USA}
\affiliation{Department of Physics, Massachusetts Institute of Technology, Cambridge, MA 02139, USA}

\date{\today}

\begin{abstract}
Magnetic exchange driven proximity effect at a magnetic insulator / topological insulator (MI/TI) interface provides a rich playground for novel phenomena as well as a way to realize low energy dissipation quantum devices. Here we report a dramatic enhancement of proximity exchange coupling in the MI/magnetic-TI EuS / Sb$_{2-x}$V$_x$Te$_3$ hybrid heterostructure, where V doping is used to drive the TI (Sb$_{2}$Te$_3$) magnetic. We observe an artificial antiferromagnetic-like structure near the MI/TI interface, which may account for the enhanced proximity coupling. The interplay between the proximity effect and doping in a hybrid heterostructure provides insights into the engineering of magnetic ordering.
\begin{description}
\item[PACS numbers]
61.05.fj, 75.25.-j, 75.30.Gw, 75.70.Cn.
\end{description}
\end{abstract}

\keywords{Interlayer Magnetic Coupling, Topological Insulator, Magnetic Proximity Effect.}
\maketitle

\begin{figure}
\centering
\includegraphics[width=1.0\columnwidth]{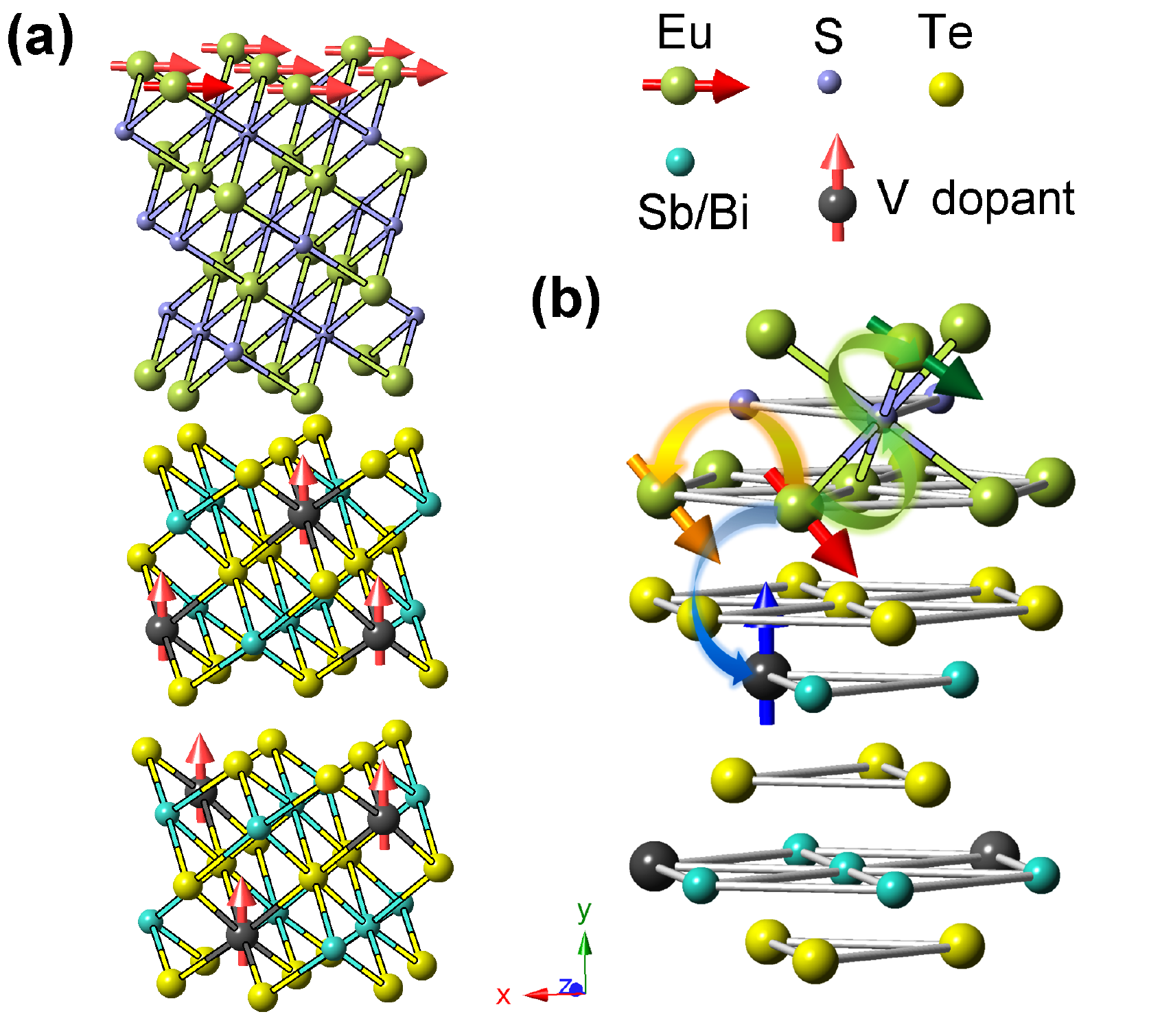}
\caption{(color). (a) MI EuS / V-doped TI Sb$_2$Te$_3$ hybrid heterostructure. The arrows denote the spin direction. The V-doped TI layer has a perpendicular magnetic anisotropy, while the magnetic EuS has in-plane magnetic anisotropy. Such heterostructure may create an exotic magnetic environment near the interface, as illustrated in (b). For a given Eu ion (red-arrow), it interacts with neighborhood intra-plane Eu (orange arrow) through Heisenberg interaction, inter-plane Eu ions (green arrow) through super-exchange interaction, spin-polarized states at TI surface and localized moments in TI (blue arrow). }
\label{Fig1}
\end{figure}

\begin{figure}
\centering
\includegraphics[width=0.97\columnwidth]{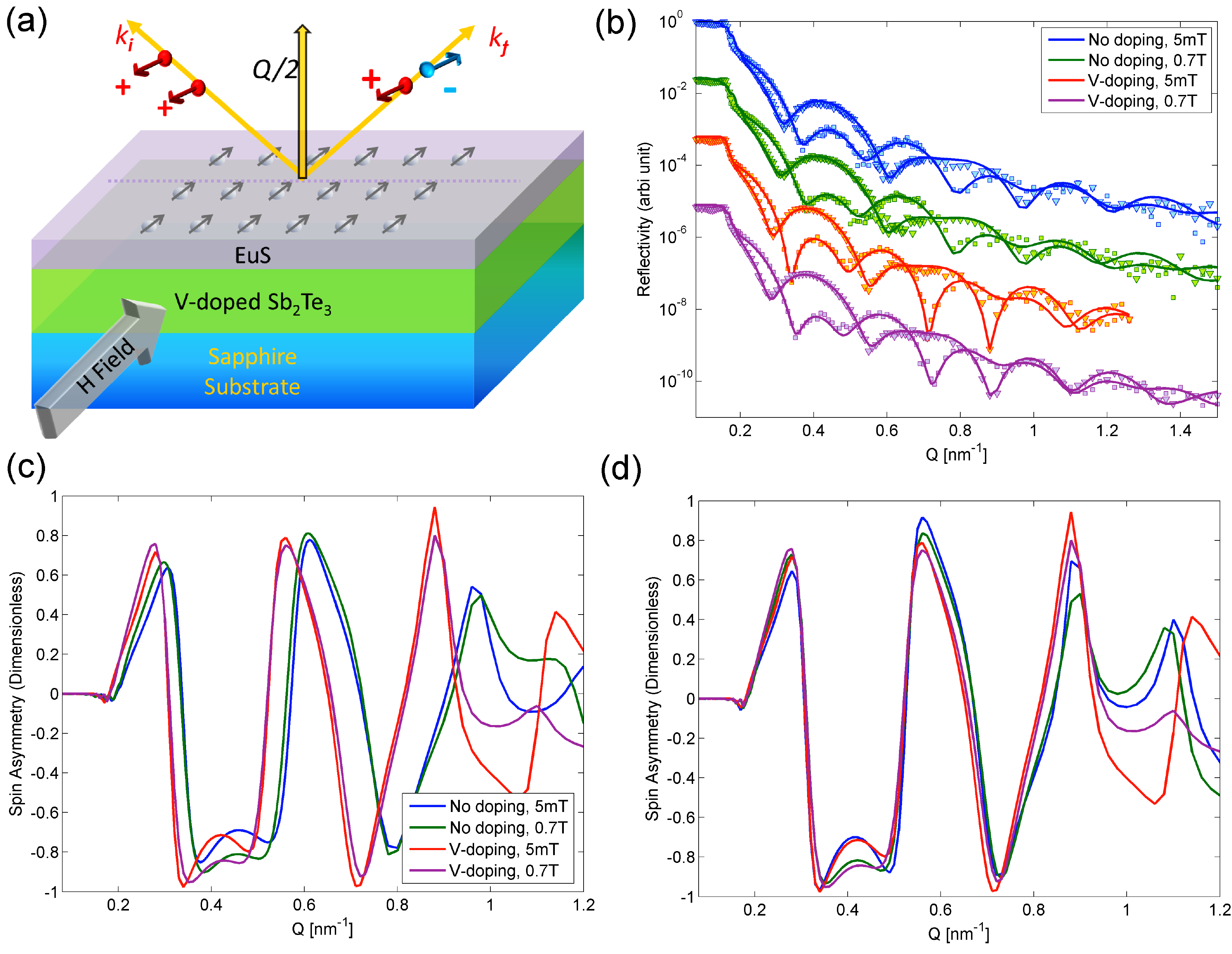}
\caption{(color). (a) The configuration of specular PNR. $k_i$, $k_f$ and $Q$ denote the incident wavevector, reflected wavevector and the momentum transfer, respectively. (b) The spin $+$ and spin $-$ PNR data $R^+$  and $R^-$ for the EuS/Sb$_{1.9}$V$_{0.1}$Te$_3$ and EuS/Sb$_2$Te$_3$ samples, at low ($\unit[5]{mT}$) and high ($\unit[0.7]{T}$) in-plane guide fields. The fitting results are represented by solid lines and the curves are shifted for clarity. (c) The spin asymmetry for the reflectivity in b. (d) The same spin asymmetry, but assuming one control sample EuS/Sb$_2$Te$_3$ have exactly the same thickness as EuS/Sb$_{1.9}$V$_{0.1}$Te$_3$. In this way, the spin asymmetry difference is dominated by only the magnetic structure instead of crystalline structure. At $\sim \unit[0.4]{ nm^{-1}}$, the difference comes from the effect of external magnetic field, while at $\sim\unit[1.0]{nm^{-1}}$ the difference mainly comes from V-dopants.}
\label{Fig2}
\end{figure}

The time-reversal symmetry (TRS) breaking and surface bandgap opening of topological insulator (TI) is an essential step towards the observation of novel quantum phases and realization for TI-based devices \cite{5qi2011topological,4hasan2010colloquium}. In general, there are two approaches to break the TRS: transitional metal (TM) ion doping, where Cr or V are doped into the entire TI \cite{6wei2013exchange,7chen2010massive,dopingPRL}, and magnetic proximity effect, where a magnetic insulator (MI) layer in proximity to TI  induces exchange coupling \cite{1lang2014proximity,3vobornik2011magnetic,6wei2013exchange,ProximityAPL}. Doping TM impurities into TI will introduce a perpendicular ferromagnetic (FM) anisotropy and provide a straightforward means to open up the bandgap of the TI’s surface state, with profound influence to its electronic structure \cite{7chen2010massive,9nature,10chang2013experimental,11xu2012hedgehog,12hor2010development,13chang2014chemical,14chang2013thin}. In particular, quantum anomalous Hall effect (QAHE), where quantum Hall plateau and dissipationless chiral edge channels emerge at zero external magnetic field, has recently been realized in Cr-doped and V-doped TIs \cite{9nature,10chang2013experimental,15qi2008topological,16yu2010quantized,17liu2008quantum,18kou2014scale,19checkelsky2014trajectory, 20chang2014breaking}. Ideally, compared to the doping method, the proximity effect has a number of advantages, including spatially uniform magnetization, better controllability of surface state, and freedom from dopant-induced scattering that degrades TI properties, as well as preserving TI intrinsic crystalline structure, etc. \cite{21PhysRevB.91.014427,22topological}. However, due to the in-plane anisotropy and low Curie temperature of the common MI, such proximity effects are usually too weak to induce strong proximity magnetism in TI. In fact, compared to magnetically doped TI which can induce as large as a $\unit[50]{meV}$ surface bandgap \cite{7chen2010massive}, the EuS/TI system has only $\unit[7] {meV}$ gap opening due to the strongly localized Eu-f orbitals \cite{23eremeev2013magnetic}. Therefore, the enhancement of proximity magnetism is highly desirable to make it a valuable approach as doping hence takes full advantage.

In this \textit{Letter}, we report significant enhancement of the proximity effect in MI EuS / magnetic-TI Sb$_{2-x}$V$_x$Te$_3$ hybrid heterostructure. Using polarized neutron reflectometry (PNR), we inferred an increase of proximity magnetization per unit cell (u.c.) in TI, from $1.2\mu_B\unit[]{/u.c}$ to $2.7\mu_B\unit[]{/u.c}$ at $x=0.1$ doping level. High-resolution transmission electron microscopy (HRTEM) image identifies the TI/EuS interfacial sharpness and excludes false positive magnetism signal from interdiffused Eu ions into TI. Furthermore, the proximity effect enhancement is accompanied by a decrease in the interfacial magnetization of EuS, resulting in an exotic antiferromagnetic (AF) like structure. The existence of the “artificial” AF structure between FM EuS and the FM Sb$_{2-x}$V$_x$Te$_3$ with different anisotropies is consistent with magnetometry measurements which shows exchange bias. Such artificial AF ordering in this FM / FM hybrid heterostructure may shed lights for designing novel magnetic phases and devices.

High-quality MI $\unit[6]{nm}$ EuS / 15 quintuple layer (QL) magnetic TI Sb$_{2-x}$V$_x$Te$_3$ hybrid heterostructures were grown by molecular beam epitaxy (MBE) under a base vacuum $\sim 5\times 10^{-10}$ Torr, where magnetic TI thin films Sb$_{2-x}$V$_x$Te$_3$ were grown on clean, heat-treated sapphire (0001) substrates with V-dopants coevaporated during TI growth. The EuS (111) layer was deposited \textit{in situ} over the TI film using electron beam source. To understand the interplay between proximity effect and TM doping, $\unit[6]{nm}$ EuS/ $\unit[15]{QL}$ Sb$_2$Te$_3$, $\unit[15]{QL}$ Sb$_{2-x}$V$_x$Te$_3$ and $\unit[15]{QL}$ Sb$_{2-x}$V$_x$Te$_3$ samples were fabricated.

The atomic configuration of the MI / magnetic TI heterostructure is shown in Fig. \ref{Fig1} (a). The upper EuS has in-plane anisotropy \cite{24kunevs2005exchange,25liu1983magnetic,26franzblau1967magnetocrystalline,27von1965ferromagnetic} within xz-plane, while the lower TM doped TI has easy axis out-of-plane \cite{14chang2013thin,20chang2014breaking,10chang2013experimental} along y-axis. The different anisotropy directions in the two layers, corroborated by a strong interfacial spin-orbit coupling, create a complex magnetic environment for the EuS near the interface (Fig. \ref{Fig1} b). The Heisenberg interaction, superexchange interaction \cite{25liu1983magnetic,29boncher2014europium}, $d-f$ coupling \cite{30lee1983magnetic} and coupling with the TI's spin texture may finally contribute to an overall augmentation of the proximity effect.

The PNR experiments were carried out using PBR beamline at the NIST Center for Neutron Research, from which the in-plane magnetization is extracted. The experimental setup is shown in Fig. \ref{Fig2}(a), where the incident spin-polarized neutron beam is reflected by the heterostructure sample, while the spin non-flip reflectivity signals from both spin components ($++$ and $- -$) were collected under external guide magnetic field. The PNR refinement is based on a multilayered TI / proximity /interfacial EuS / main EuS model \cite{22topological}. To maximize the PNR information extraction, we did not compare the $\chi^2$ with and without proximity effect due to limited sensitivity. Instead, we presume the existence of the proximity coupling layer and optimize its magnitude. 

The spin non-flip reflectivity curves for the MI/ magnetic TI sample EuS/Sb$_{1.9}$V$_{0.1}$Te$_3$ and control sample MI/TI EuS/Sb$_2$Te$_3$, at low ($\unit[5]{mT}$) and high ($\unit[0.7]{T}$) fields, are shown in Fig. \ref{Fig2} (b). The fitting and refinement of PNR is performed using the GenX program \cite{31bjorck2007genx}. To directly infer the possible contribution of V-dopants, the corresponding spin asymmetries $SA=\tfrac{R^+-R^-}{R^++R^-}$  are plotted (Fig. \ref{Fig2}c) for the raw data and thickness-adjusted data (Fig. \ref{Fig2}d). In this way, the different features of the $SA$ in Fig. \ref{Fig2} (d) are solely coming from the magnetic structure since the crystalline structure is taken to be identical. We see that at $Q\sim \unit[0.4]{ nm^{-1}}$, $\mu_{0}H=\unit[5]{mT}$ SA for both samples with and without V-dopants overlap each other, but distinct with the $\mu_{0}H=\unit[700]{mT}$ $SA$ curves, indicating an effect from guide field within this Q range; while at $Q\sim\unit[1.0]{nm^{-1}}$, a splitting of the $SA$ curves for both samples at same guide field (eg. blue and red curves) is observed. This indicates the influence of the V-dopants to magnetic structure at high Q range (spatially localized) even without performing fitting.

\begin{figure}
\centering
\includegraphics[width=0.97\columnwidth]{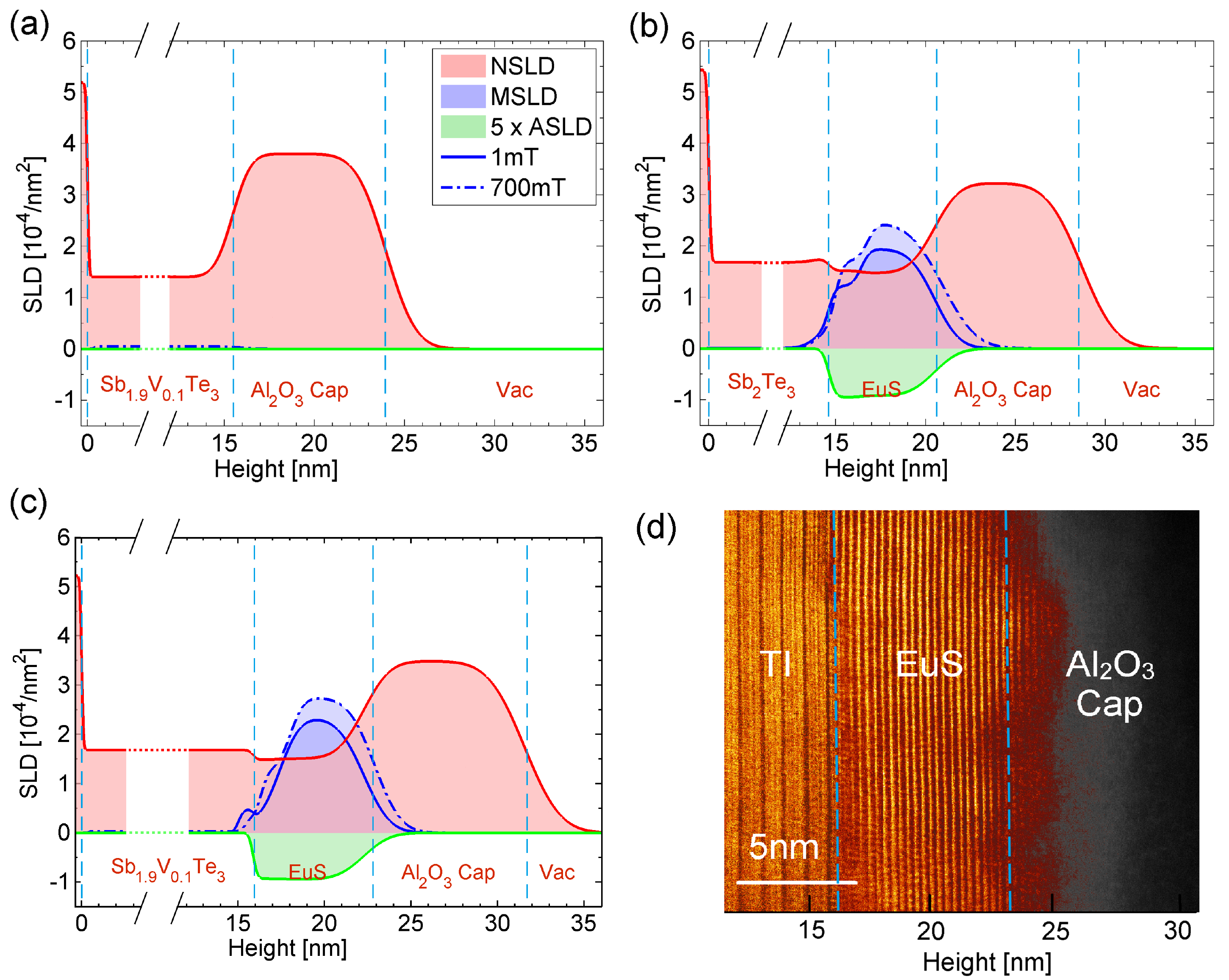}
\caption{(color). PNR fitting profiles of doping-only sample Sb$_{1.9}$V$_{0.1}$Te$_3$ (a), proximity only sample EuS / Sb$_2$Te$_3$ (b) and hybrid heterostructure sample EuS / Sb$_{1.9}$V$_{0.1}$Te$_3$ (c). The NSLD, MSLD and ASLD denote the nuclear, magnetic and absorption scattering length density, respectively. The proximity effect appears as finite magnetization signal (blue curves) in the region of TI near the interface ($\sim\unit[15]{nm}$). The absorption free feature in this region excludes the possible contribution of interdiffused Eu ions. We see clearly that with V-doping, the proximity magnetism is enhanced as a bump in (c), accompanied with a further suppression of magnetism of interfacial EuS ($\unit[15-18]{nm}$). (d)HAADF TEM image of the EuS / Sb$_{1.9}$V$_{0.1}$Te$_3$ hybrid heterostructure. A sharp interface between the TI / MI is developed, indicating an epitaxial growth of EuS, consistent with (c) for uniformly distributed ALSD of EuS. The island-like crystalline facets between EuS and Al$_2$O$_3$ cap explains the roughness in (c).}
\label{Fig3}
\end{figure}

\begin{figure}
\centering
\includegraphics[width=0.97\columnwidth]{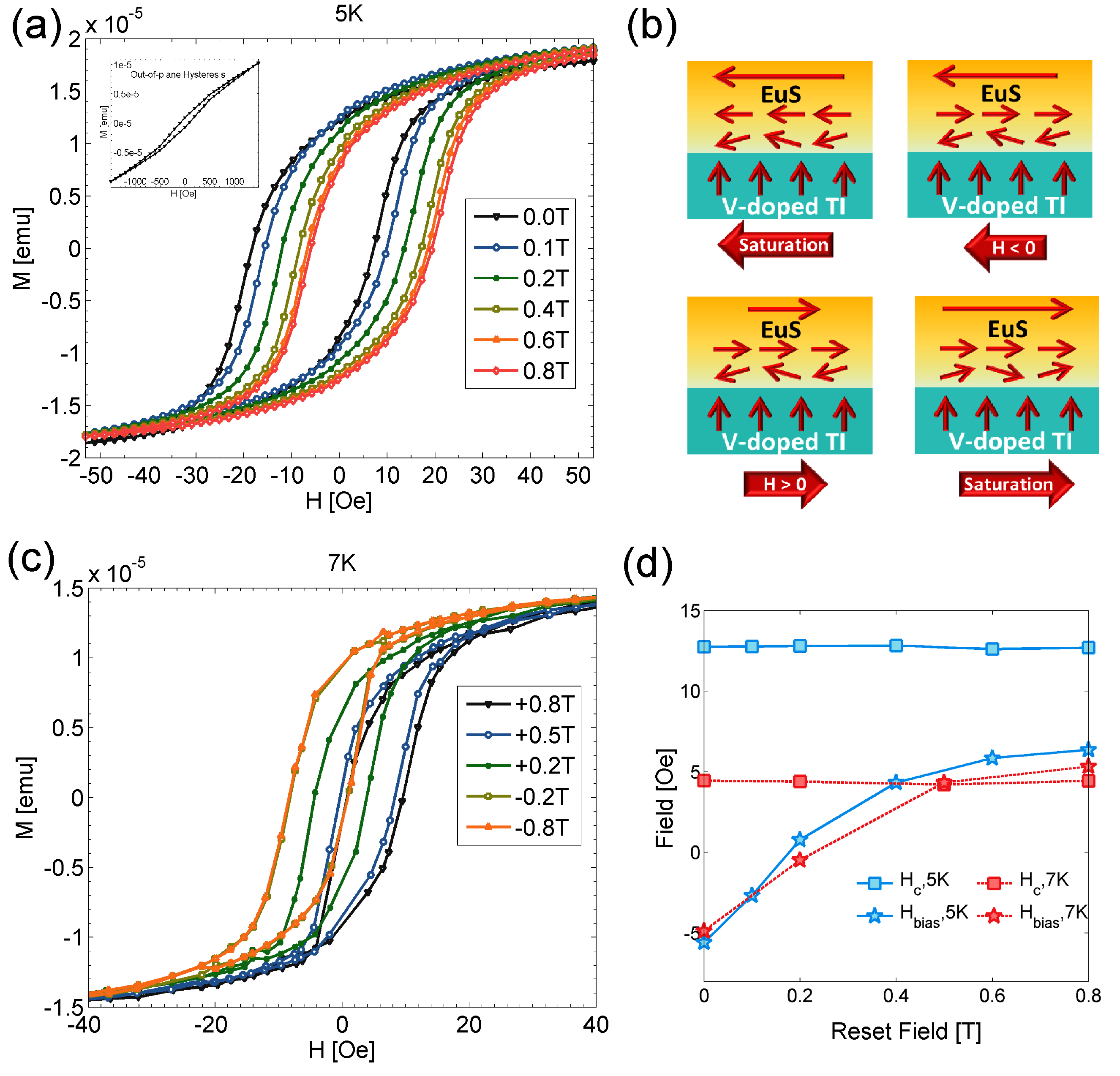}
\caption{(color). Magnetic measurements of a 2 nm EuS /10 QL Bi$_1.9$V$_0.1$Te$_3$ hybrid heterostructure using a Quantum Design SQUID magnetometer. (a)In-plane hysteresis showing that a negative EB following a set field of -1T, can be switched to positive bias by applying a positive resetting field. Inset is the out-of-plane magnetic hysteresis of the same sample, showing a finite remanent moment. (b, c) EB and coercive field as a function of the in-plane resetting field at 5K and 7K, respectively. (d) Illustration of EB where the moments at the interface of EuS are pinned by the exchange-coupled moments in the V-doped TI.}
\label{Fig4}
\end{figure}

Results of fitting the PNR data are shown in Fig. 3. The substrate lies in the region below $\unit[0]{nm}$. Nuclear-SLD (NSLD, red curves) identifies the chemical compositional contrast, where the NSLDs for each compound layer are correctly reproduced from PNR fitting (sapphire substrate $\unit[5.5\times 10^{-5}]{nm^{-2}}$ , Sb$_2$Te$_3$ $\unit[1.8\times 10^{-4}]{nm^{-2}}$ , EuS $\unit[1.5\times 10^{-4}]{nm^{-2}}$ and amorphous Al$_2$O$_3$ capping layer $\unit[4\times 10^{-4}]{nm^{-2}}$). This further validates the fitting quality. In Fig. \ref{Fig3}(a), without the EuS proximity layer, the V-dopants in the Sb$_{1.9}$V$_{0.1}$Te$_3$ sample contribute to no more than $0.2\ub$  in-plane magnetization at $\mu_{0}H=\unit[0.7]{T}$, indicating a very strong perpendicular FM anisotropy in V-doped TI. This is consistent with the result in the inset of Fig. \ref{Fig4} (a), and facilitates us in obtaining reliable PNR refinement by fixing the magnetization of the magnetic TI layer. The magnetic-SLD (MSLD) (blue curves) at $\mu_{0}H$ = 5mT and 700mT in-plane guide fields at T = 5 K are also plotted. In Fig. \ref{Fig3} (b), for the EuS/pure TI sample, we see a penetration of magnetization into TI. Unlike the EuS region where strong absorption LSD (ASLD) is always accompanied due to the Eu ions' large neutron absorption, the magnetization into TI does not show any absorption ($\sim\unit[14-15]{nm}$), indicating that such magnetism in TI is not from ferromagnetic Eu ions interdiffused into Sb$_2$Te$_3$, but from proximity exchange effect. The free of interdiffusion is also consistent with our Transmission Electron Microscopy (TEM) result in Fig. \ref{Fig3}(d), where a sharp interface between epitaxially grown EuS and Sb$_{1.9}$V$_{0.1}$Te$_3$ is developed.

The magnetization at the interface in proximity structures is greatly enhanced when V-dopant is at present, from $1.2\ub$ (Fig. \ref{Fig3}b, without V-doping) to $2.7\ub$ (Fig. \ref{Fig3}c, V-doped), i.e. almost tripled. In both cases, the penetration depth of proximity is $\sim\unit[1]{nm}$, consistent with Bi$_2$Se$_3$ / EuS interface \cite{22topological}. Besides, the in-plane magnetization of EuS drops dramatically near the interface, from $\sim 3\ub$ without V-dopants to $\sim 0.5\ub$ with V-doping, at $\mu_{0}H=\unit[5]{mT}$. This is due purely to magnetic effect instead of interfacial roughness since the ASLD is flat near the TI interface. On the contrary, magnetization drop at EuS / Al$_2$O$_3$ interface ($\unit[\sim23]{nm}$) is due to the Stranski–Krastanov growth \cite{growth}, leading to a thickness variation and formation of island. This is directly confirmed from Z-contrast high-angle angular dark field (HAADF) TEM image (Fig. \ref{Fig3}(d)). At higher field $\mu_{0}H=\unit[0.7]{T}$, an increase of the in-plane EuS magnetism is accompanied with a drop of proximity effect into TI. Since only the perpendicular direction magnetism will contribute to the proximity effect \cite{5qi2011topological}, a high in-plane guide field tends to align the EuS moment in-plane and reduce the proximity.

To understand the origin of the in-plane EuS magnetism drop near the interface, we examined the exchange bias (EB) of the magnetic hysteresis measurements. Fig. \ref{Fig4} plots the results of low-field in-plane hysteresis measurements of a 2nm EuS / 10QL Bi$_1.9$V$_0.1$Te$_3$ hybrid heterostructure instead of Sb$_2$Te$_3$ since both belong to Bi$_2$Se$_3$ TI family and share very similar crystalline structure. Despite Sb$_2$Te$_3$ is more suitable for PNR studies due to less interstitial V-defects, Bi$_2$Te$_3$ is more suitable for SQUID due to higher diamagnetic susceptibility. Fig. \ref{Fig4}(a) shows that the EB can be switched from negative to positive by a field of $\mu_{0}H=\pm1$T. Fig. \ref{Fig4}(b, c) show the characteristics of the EB as a function of the resetting field, at 5K and 7K, respectively. We adopt the traditional approach for exchange bias (EB) measurement \cite{32kiwi2001exchange,33stamps2000mechanisms,34nogues1999exchange} at various reset fields, where the EB was initially set negative by applying a field of $-1$T, which was followed by a positive resetting field and then measuring the low-field hysteresis \cite{35wen2014asymmetric}. This was repeated for resetting fields from 0 to + 0.8 T, where the bias is seen to shift from $H_{bias} = –5$ to $+6$Oe. Results of the exchange biasing strongly suggests the existence of an AF structure and a likely magnetic frustration \cite{36schlenker1986magnetic}, and is quite striking since our system is only composed by two strong FMs. The possible configuration derived from EB is illustrated in Fig. \ref{Fig4}(d), where V-doped TI keeps a perpendicular anisotropy, but an interfacial AF structure is created to cause the EB.

To further understand the implication of the results in Fig. \ref{Fig4}, we develop a phenomenological energy model to describe the coupling between the FM and AF. The anisotropic energy for bulk EuS can be written as \cite{26franzblau1967magnetocrystalline}
\begin{equation}\label{eq1}
{{E}_{an}}\!=\!{{\kappa }_{1}}{{M}_{s}}t\left( \alpha _{1}^{2}\alpha _{2}^{2}\!+\!\alpha _{1}^{2}\alpha _{3}^{2}\!+\!\alpha _{3}^{2}\alpha _{2}^{2} \right)\!+\!{{\kappa }_{2}}{{M}_{s}}t\alpha _{1}^{2}\alpha _{2}^{2}\alpha _{3}^{2}
\end{equation}
where $\alpha_i$ is the directional cosine along i$^{th}$ direction, $M_s$ is the saturation magnetism per area, $t$ is the thickness of FM layer and the anisotropic constants $\kappa_1=\unit[-19.6]{Oe}$  and $\kappa_2=\unit[-4.6]{Oe}$  at $T=\unit[1.3]{K}$ \cite{26franzblau1967magnetocrystalline}. Since our interest is in thin film structures  with a single symmetry axis ($y$-axis in Fig. \ref{Fig1} a), eq. (\ref{eq1}) could be rewritten as using a simplified model for hexagonal and cubic lattice \cite{37skomski2008simple},
\begin{equation}\label{eq2}
E_{an}=K_1M_st \sin^2(\theta)+K_2M_st \sin^4(\theta)
\end{equation}
where $K_1=\kappa_1=\unit[-19.6]{Oe}$ , $K_2=-\frac{7}{8}\kappa_1+\frac{1}{8}\kappa_2=\unit[16.6]{Oe}$, $\theta$ is the angle between the magnetization and the symmetry axis. Since $K_1<0$, $\theta=\pi/2$ corresponds in the present case for EuS showing easy-plane anisotropy within $xz$-plane. For a thin film, we further define $K_{1,eff}=K_s/t+K_1$. We require the surface anisotropy constant $K_s>0$, since for thinner sample $K_{1,eff}$  will approach zero from the negative side, indicating a rotation of the in-plane easy plane to an out-of-plane direction, resulting in a magnetic canting which is reasonable for compensated thin film interfaces \cite{33stamps2000mechanisms}.

Taking into account the external magnetic field $H$ and the FM/AF coupling $J$, the total energy could be written as
\begin{eqnarray}\label{eq3}
   E&=&-H{{M}_{s}}t\sin (\theta) -J\sin (\theta) \nonumber \\
  &+&\left( {{K}_{\text{1}}}+\frac{{{K}_{s}}}{t} \right){{M}_{s}}t{{\sin }^{2}}(\theta) +{{K}_{\text{2}}}{{M}_{s}}t{{\sin }^{4}}(\theta)
\end{eqnarray}
At saturation field configuration $\theta=\pm \pi/2$ and considering the energy extreme, we obtain the bias field and coercivity
\begin{equation}\label{eq5}
H_{bias}=-\frac{J}{Mt},\quad H_c=2K_1+4K_2+\frac{2K_s}{t}
\end{equation}
respectively.
\setlength{\extrarowheight}{9pt}
\begin{table}
\caption{Temperature dependence of the anisotropy constants. Green, Blue, red and black colored values are from \cite{26franzblau1967magnetocrystalline}, eq. (\ref{eq6}), measurements in Fig.\ref{Fig4} and eq. (\ref{eq5}), respectively.}\label{tab:1}
\begin{tabularx}{0.47\textwidth}{  X  X X  X }
    \hline
    $T_c=\unit[17]{K}$ & $\unit[1.3]{K}$ & $\unit[5]{K}$ & $\unit[7]{K}$ \\ \hline
    $K_1(\unit[]{Oe})$ & \textcolor{ForestGreen}{-19.6} & \textcolor{blue}{-13.10} & \textcolor{blue}{-9.96} \\ \hline
    $K_2(\unit[]{Oe})$ & \textcolor{ForestGreen}{+16.6} & \textcolor{blue}{7.41} & \textcolor{blue}{4.29} \\ \hline
   $H_c(\unit[]{Oe})$ & N/A & \textcolor{red}{12.7} & \textcolor{red}{4.4} \\
    \hline
     $K_s/t(\unit[]{Oe})$ & \textcolor{blue}{13.8} & 9.3 & 7.2; \textcolor{blue}{7.1} \\
     \hline
\end{tabularx}
\end{table}

The anisotropic coefficients strongly depend on temperature \cite{37skomski2008simple}. In the mean-field approximation, the temperature dependence of anisotropy can be expressed using the Callen and Callen theory as \cite{39skomski2006finite}
\begin{equation}\label{eq6}
K(T)=K(0){{\left( 1-\frac{T}{{{T}_{c}}} \right)}^{\frac{n(n+1)}{4}}}
\end{equation}
where $n$ is the order of anisotropy constant, $n(K_1)=2$  and $n(K_2)=4$. Assuming that the Curie temperature of EuS is $T_c=\unit[17]{K}$, we obtain the temperature dependence of anisotropy constants as shown in Table \ref{tab:1}. One remarkable feature for this model is that the surface anisotropy $K_s/t$  calculated from experimental values and eq. (\ref{eq5}) coincides with the independent check using eq. (\ref{eq6}), giving $\unit[7.2]{Oe}$ vs $\unit[7.1]{ Oe}$ at $\unit[7]{K}$. Finally, this yields a surface anisotropy $K_s=\unit[0.0014]{erg\cdot cm^{-2}}$  by assuming a $\unit[2.5\times 10^{-5}]{emu}$  saturation and $\unit[5]{mm^2}$ sample area. This term is the origin of magnetic canting of interfacial EuS.

Contrary to the strong $T$-dependence of anisotropy, the bias field $H_{bias}$ thus AF/FM coupling constant $J$ has a weak dependence with temperature, indicating an origin of FM/AF coupling different from magnetic crystalline anisotropym such as the prominent role of Spin-Orbit interaction and spin-momentum locking at the surface of the TI.

To summarize, we have reported a large enhancement of proximity exchange coupling strength in MI/magnetic-TI hybrid heterostructure. This overcomes the major disadvantage in MI/TI heterostructures where the proximity coupling strength is considered weak \cite{23eremeev2013magnetic}. To our knowledge, this is also the first report combining TM doping and proximity effect to magnetize TIs. Here, the magnetic dopants magnetize the TI surface states, which are further coupled to the MI; whereas the magnetic TI has a strong perpendicular anisotropy that compensates the weakness of MI with in-plane anisotropy. The reduction of interfacial magnetism is consistent with the exchange bias result, where an AFM-like structure is artificially created, where a $K_s=\unit[0.0014]{erg\cdot cm^{-2}}$ surface anisotropy is extracted. Despite this value being small compared to the stronger examples such as the Au/Co interface \cite{41bruno1988magnetic}, this approach provides fruitful insights to tailor new magnetic structure at TI/MI interfaces.

\acknowledgements
M. L. would thank the helpful discussion with Prof. Albert Fert. J.S.M. and C.Z.C would thank support from the STC Center for Integrated Quantum Materials under NSF grant DMR-1231319, NSF DMR grants 1207469 and ONR grant N00014-13-1-0301. M. J. and D. H. acknowledge support from NSF DMR-907007 and NSF ECCS-1402738. L.W. and Y.Z. were supported by DOE-BES under Contract No. DE-AC02-98CH10886.
\bibliography{EnhancedProximity}
\end{document}